\documentclass[aps,prc,preprint,groupedaddress]{revtex4-1}
\usepackage{amsmath}
\usepackage{graphicx}
\usepackage{bm}
\begin{document}
\title{Hartree-Fock Gamow basis from realistic nuclear forces}
\author{Q. Wu}
\author{F.R. Xu}
\email{frxu@pku.edu.cn}
\affiliation{School of Physics, and State Key Laboratory of Nuclear Physics and Technology, Peking University, Beijing 100871, China}
\begin{abstract}
  We present a simplified method to generate the Hartree-Fock Gamow basis from realistic nuclear forces. The Hartree-Fock iteration in the harmonic-oscillator basis is first performed, and then the obtained HF potential is analytically continued to the complex-k plane, finally by solving the Schr\"odinger equation in the complex-k plane the Gamow basis is obtained. As examples, the method is applied to $^4$He and $^{22}$O with the renormalized chiral N$^3$LO potential. The basis obtained which includes bound, resonant and scattering states can be further used in many-body calculations to study weakly bound nuclei.
\end{abstract}
\maketitle
\section{Introduction}
  The description of nuclei far from the valley of stability is a challenge in nuclear physics. These nuclei exhibit unusual properties such as halo densities and weak bindings. The coupling to the continuum is needed for the theoretical description of these exotic nuclei. The Gamow shell model \cite{ref1,ref2,ref3,ref4,ref5,ref6,ref7,ref8,ref9,ref10} which extends the traditional shell model to the complex plane, can efficiently include the continuum coupling. This model unifies nuclear structure and reaction properties, and has proven to be a promising tool for the descriptions of weakly bound and unbound nuclei. The starting point of the Gamow shell model is the Berggren completeness where bound, resonant (or Gamow) and scattering states are treated on an equal footing \cite{ref11,ref12,ref13,ref14,ref15}. Employing the Berggren basis, first principle calculations can also be extended to study weak-binding nuclei and nuclear reactions.

  A big concern is how to obtain the single-particle Berggren basis which is used to construct the Slater determinants in many-body calculations. The single-particle basis in Gamow shell-model calculations are normally constructed from a Woods-Saxon potential with parameters fitted to experimental single-particle energies \cite{ref1,ref10}. However, in a fully microscopic approach, the single-particle basis constructed from the nucleon-nucleon interaction is preferred. In many {\it ab initio} calculations, the Hartree-Fock (HF) iteration is usually performed as a first approximation \cite{ref16,ref17,ref18}. When extending the {\it ab initio} calculations to handle resonant states employing the Berggren ensemble, it's very useful to have the single-particle Berggren basis generated from the HF potential.

  There have been several studies about how to obtain the Berggren basis. In Refs. \cite{ref19,ref20}, mature numerical methods have been developed to solve the complex eigenvalue problem of the Schr\"odinger equation with a local potential. For a nonlocal HF potential, the numerical procedure known as the Hartree-Fock Gamow (GHF) method is adopted in Refs. \cite{ref21,ref1,ref22} to generate a basis that includes resonant and nonresonant states. In all these calculations, the basis is obtained in the coordinate space by taking proper boundary conditions. In Ref. \cite{ref23}, however, the derivation of the Gamow HF basis in momentum space is carried out. In momentum space, there is no need to worry about the boundary conditions.
  
  In this paper, we present an alternative method to generate the Gamow HF basis from realistic nucleon-nucleon interactions. As in Ref. \cite{ref23}, the basis in our method is finally obtained with the contour deformation method in the complex momentum space \cite{ref9}, while we adopt a simplified way to get the HF potential in the complex $k$-plane. As examples, we apply this method to $^4$He and $^{22}$O with the chiral two-body N$^3$LO potential \cite{ref24} softened by V$_{\text{low-}k}$ \cite{ref25}.


\section{Method}
The HF approximation is a common many-body method where the wave function of the system is described by a single Slater determinant. This method is also called self-consistent mean field approximation since the solution to the HF equation behaves as if each particle of the system is subjected to the mean field created by all other particles. The mean field which is called HF potential is usually obtained by solving the nonlinear HF equation by iteration.

The HF calculations for nuclei starting from realistic nuclear forces have been successful \cite{ref17}. The $A$-body intrinsic nuclear Hamiltonian $H$ is
\begin{align}
    H=& \frac{\hbar^2}{2mA}\sum\limits_{i<j}^A(\bm{k}_i-\bm{k}_j)^2 + \sum\limits_{i<j}^A V_{ij}\nonumber\\
    =&\left (1-\frac{1}{A}\right )\sum\limits_{i=1}^A \frac{\hbar^2\bm{k}_i^2}{2m} + \sum\limits_{i<j}^A \left ( V_{ij}-\frac{\hbar^2}{mA} \bm{k}_i\cdot\bm{k}_j\right ),
    \label{eq:h}
\end{align}
where $m$ is the mass of the nucleon, $V_{ij}$ is the two-body nucleon-nucleon force.
With the Hamiltonian (\ref{eq:h}), the symmetry-preserved HF iteration is first performed in the harmonic-oscillator (HO) basis \cite{ref17}. For open-shell nuclei, the average filling can be adopted as in Ref. \cite{ref1}. We only deal with the situation where the single-particle states below the Fermi surface are bound, which is true in most cases. 
When the iteration converges, the HF potential in HO basis denoted as $\langle n_1l|U_{\text{HF}}^{(l,j,t_z)}|n_2l \rangle$ is established, where $(l,j,t_z)$ is used to label the channel with orbital angular momentum $l$, total angular momentum $j$ and isospin $t_z$. $|nl\rangle$ is the HO state. 
The HF potential is generally nonlocal in r-space, thus the use of the shooting method \cite{ref20} is not straightforward.

It is preferred  to solve the one-body Schr\"odinger equation in momentum space.
The HF equation in momentum space can be written formally as
\begin{equation}
 \frac{\hbar^2}{2\mu}k^2 \psi_{nljt_z}(k) + \int dk' k'^2 \langle k|U_{\text{HF}}^{(l,j,t_z)}|k'\rangle \psi_{nljt_z}(k')=E_{nljt_z} \psi_{nljt_z}(k),
\label{eq:1}
\end{equation}
where we have introduced the effective mass $\mu=(1-1/A)^{-1}m$. $\psi_{nljt_z}$ is the HF single-particle state in $(l,j,t_z)$ channel. There are several advantages solving the equation in momentum space. First, the boundary conditions are automatically built into the integral equation. Secondly, the momentum representaion of the Gamow states are non-oscillating and rapidly decreasing, as opposed to the coordinate representation. The numerical procedures are often easier to implement.

The Gamow (or resonant) states are generalized eigenstates of the Schr\"odinger equation with complex energy eigenvalues $E=E_0-i \Gamma/2$, where $\Gamma$ stands for the decay width. These states correspond to the poles of the $S$ matrix in the complex energy plane lying below the positive real axis \cite{ref2}.
To obtain the resonant states, Eq. (\ref{eq:1}) needs to be generalized to the complex $k$-plane. This can be done by replacing the integral path in Eq. (\ref{eq:1}) from the real axis to a contour in the complex $k$-plane \cite{ref9}. Then the Schr\"odinger equation reads
\begin{equation}
  \frac{\hbar^2}{2\mu}k^2 \psi_{nljt_z}(k) + \int_{L^+} dk' k'^2 \langle k|U_{\text{HF}}^{(l,j,t_z)}|k'\rangle \psi_{nljt_z}(k')=E_n \psi_{nljt_z}(k),
\label{eq:2}
\end{equation}
where $L^+$ in a contour in the lower half complex $k$-plane.

The key problem now is how to obtain the HF potential $U_{\text{HF}}$ in the complex $k$-plane. In Ref. \cite{ref23}, the double Fourier-Bessel transformation is adopted to achieve the analytical continuation from the real $k$ axis to the complex $k$-plane. Here, we employ a simplified approach where the direct basis transformation is used.
\begin{equation}
 \langle k|U_{\text{HF}}^{(l,j,t_z)}|k'\rangle = \langle k|n_1l\rangle \langle n_1l|U_{\text{HF}}^{(l,j,t_z)}|n_2l\rangle \langle n_2l|k'\rangle,
\label{eq:3}
\end{equation}
where $\langle k|nl\rangle$ is the momentum-space radial wave function of the HO state. $\langle k|nl\rangle$ reads
\begin{equation}
\langle k|nl\rangle=(-i)^{2n+l} e^{-1/2b^2k^2} (bk)^l \sqrt{\frac{2n!b^3}{\Gamma(n+l+3/2)}} L_n^{l+1/2}(b^2k^2),
\label{eq:4}
\end{equation}
where $L_n^{l+1/2}$ are the generalized Laguerre polynomials, $b$ is the HO length and is related to the HO frequency $\omega$ by $b=\sqrt{\hbar/m\omega}$. Since the analytical continuation of the generalized Laguerre polynomial is straightforward, the continuation of $U_{\text{HF}}$ to the complex $k$-plane can be achieved by Eq. (\ref{eq:3}).

The momentum-space Schr\"odinger equation (\ref{eq:2}) is solved by discretizing the integral interval by the Gauss-Legendre quadrature. The discretized equation reads
\begin{equation}
\frac{\hbar^2}{2\mu}k_\alpha^2 \psi_{nljt_z}(k_\alpha)+\sum\limits_\beta \omega_\beta k_\beta^2 \langle k_\alpha | U_{\text{HF}}^{(l,j,t_z)} | k_\beta \rangle \psi_{nljt_z}(k_\beta) =E_n\psi_{nljt_z}(k_\alpha),
\label{eq:5}
\end{equation}
where $k_\beta$ are the integration points and $\omega_\beta$ are the corresponding quadrature weights. By introducing $\bar{\psi}_{nljt_z}(k_\alpha)=\psi_{nljt_z}(k_\alpha) k_\alpha \sqrt{\omega_\alpha}$, Eq. (\ref{eq:5}) becomes a matrix eigenvalue problem
\begin{equation}
 \sum\limits_\beta h_{\alpha\beta} \bar{\psi}_{nljt_z}(k_\beta)=E_n \bar{\psi}_{nljt_z}(k_\alpha),
\end{equation}
where the matrix elements are
\begin{equation}
  h_{\alpha\beta}=\frac{\hbar^2}{2\mu}k_\alpha^2 \delta_{\alpha\beta}+\sqrt{\omega_\alpha\omega_\beta}k_\alpha k_\beta \langle k_\alpha | U_{\text{HF}}^{(l,j,t_z)} | k_\beta \rangle.
  \label{eq:6}
\end{equation}
The Gamow HF basis which includes bound, resonant and scattering states can be obtained by diagonalizing the complex symmetric matrix (\ref{eq:6}). The obtained basis can be written in coordinate space as 
  \begin{align}
    \psi_{nljt_z}(r)&=\int_0^\infty \psi_{nljt_z}(k) j_l(kr) k^2 dk\nonumber\\
    &=\sum\limits_\alpha k_\alpha\sqrt{\omega_\alpha} j_l(k_\alpha r) \bar{\psi}_{nljt_z}(k_\alpha).
    \label{eq:7}
  \end{align}


\section{Calculations and discussions}
To test our method, the single-particle Gamow HF states of the closed-shell $^4$He and $^{22}$O are calculated with the chiral N$^3$LO inteaction \cite{ref24}. The interaction is renormalized by $V_{\text{low-}k}$ method \cite{ref25} with a cutoff parameter $\Lambda=2.1$ fm$^{-1}$. In all the calculations, we take the frequency parameter $\hbar\omega=22$ MeV for the underlying HO basis and truncate the basis with $N_{\text{max}}=2n+l=10$.

\subsection{Resonant states}
\begin{figure}[]
\centering
\includegraphics[scale=1.0]{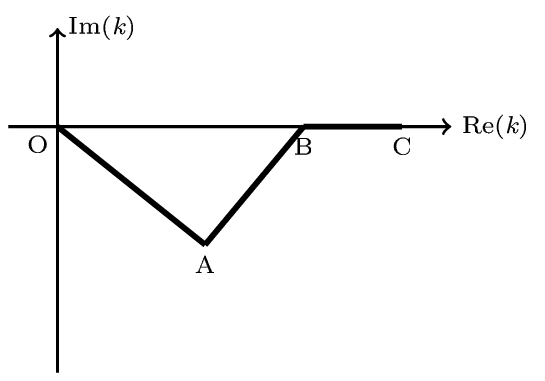}
\caption{Contour $L^+$ in the complex $k$-plane used in our calculations of resonant states. The contour is made up of three segments defined by three points A, B and C.}
\label{fig:contour}
\end{figure}

\begin{figure}[]
\centering
\includegraphics[scale=1.0]{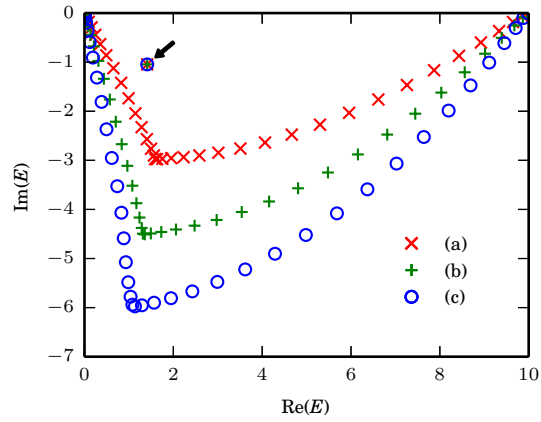}
\caption{Obtained neutron single-particle energies (in MeV) of the Gamow HF basis in the $p_{3/2}$ partial wave for $^4$He with different contours. The arrow indicates the resonant sate, whose position is stable with respect to changes of the contour. The three contours used are detailed in the text.}
\label{fig:He4_p3_2}
\end{figure}

\begin{figure}[]
\centering
\includegraphics[scale=1.0]{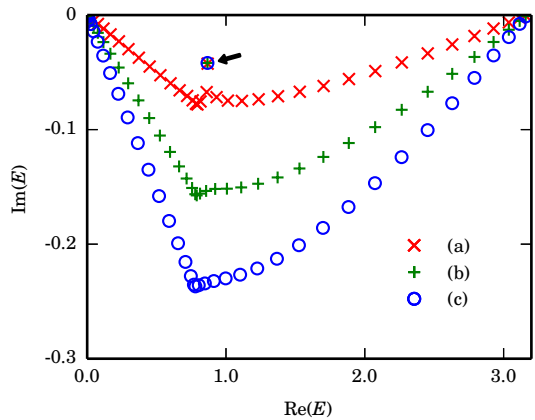}
\caption{Same as in Fig. \ref{fig:He4_p3_2} but for $^{22}$O in the $d_{3/2}$ partial wave.}
\label{fig:O22_d3_2}
\end{figure}

For $^4$He, only the lowest neutron and proton $s_{1/2}$ single-particle states in our HF calculation are bound. Resonant states may emerge in the $p_{3/2}$ partial wave. To obtain the possible resonant states, we employ the contour in the complex $k$-plane as shown in Fig. \ref{fig:contour}. The resonant state can be identified by changing the contour, since the position of the resonant state is stable with respect to changes of the contour. In Fig. \ref{fig:He4_p3_2}, we present the calculated neutron single-particle energies in the $p_{3/2}$ partial wave for $^4$He with different contours. A resonant state with an energy $E=1.412-1.046 i$ (in MeV) is clearly found. In the calculations, the adopted three contours in the complex-$k$ plane are: (a) $0\rightarrow 0.4-0.24 i\rightarrow 0.8 \rightarrow 4, (b) 0\rightarrow 0.44-0.33 i\rightarrow 0.8 \rightarrow 4$, and (c) $0\rightarrow 0.48-0.4 i\rightarrow 0.8\rightarrow 4$ (all in fm$^{-1}$). 20 points are taken for the Gauss-Legendre quadrature in each segments of the contour.

In Fig. \ref{fig:O22_d3_2}, we present the obtained neutron single-particle energies in $d_{3/2}$ partial wave for $^{22}$O with three different contours in complex-$k$ plane. The three contours are: (a) $0\rightarrow 0.2-0.01 i\rightarrow 0.4 \rightarrow 4, (b) 0\rightarrow 0.2-0.02 i\rightarrow 0.4 \rightarrow 4$, and (c) $0\rightarrow 0.2-0.03 i\rightarrow 0.4\rightarrow 4$ (all in fm$^{-1}$), see Fig. \ref{fig:contour}. The energy of the obtained resonant state indicated by the arrow in Fig. \ref{fig:O22_d3_2} is $E=0.867-0.042i$ (in MeV). In the calculations, we take 20 discretization points in each segments of the contour.

\subsection{Convergence}
We check the numerical convergence of the resonant state with respect to the number of integration points. In Table \ref{tab:O22_convergence}, we display the obtained energies of the $d_{3/2}$ resonant state in the $^{22}$O calculations with different numbers of discretization points. The contour $L^+$ employed in the calculations is $0\rightarrow 0.2-0.02 i\rightarrow 0.4 \rightarrow 4$. We can see the discretization of $L^+$ with 20 points in each segment yields a precision of the energy calculation better than 0.1 KeV for the resonant state.

\begin{table}[]
\caption{\label{tab:O22_convergence} The obtained energies (in MeV) of the $d_{3/2}$ resonant state in the $^{22}$O calculations with different numbers of discretization points. The contour used is detailed in the text. $N_{OA}$, $N_{AB}$ and $N_{BC}$ denote the number of points in the segments $OA$, $AB$ and $BC$, respectively, see Fig. \ref{fig:contour}.}
\doublerulesep 0.1pt \tabcolsep 13pt 
\begin{tabular}{ccccc}
\toprule
$N_{OA}$ & $N_{AB}$ & $N_{BC}$ & Re(E) & Im(E)\\\hline
5   &   5  & 20 &  0.8708  &  -0.0446\\
10  &  10 &  20 &  0.8666  &  -0.0421\\
20  &  20 &  20 &  0.8666  &  -0.0418\\
25  &  25 &  20 &  0.8666  &  -0.0418\\
30  &  30 &  30 &  0.8666  &  -0.0418\\
\hline
\end{tabular}
\end{table}

\subsection{Phase shifts from scattering states}
\begin{figure}[]
\centering
\includegraphics[scale=1.0]{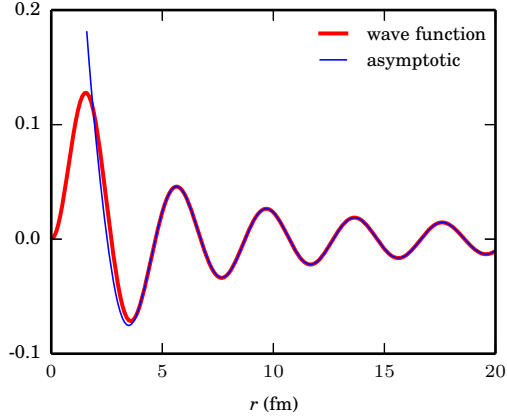}
\caption{The wave function of the calculated $^{22}$O HF single-particle scattering state at $E=50.7$ MeV as well as the asymptotic wave function in $d_{3/2}$ neutron channel.}
\label{fig:wavefun}
\end{figure}

If we employ the contour on the real axis, we can only obtain scattering sates with real energies. However, the information of the resonance states can still be extracted from the scattering states. The asymptotic behaviour of the scattering state $\psi_{nl}$ with energy $E_n$ is
\begin{equation}
  \psi_{nl}(r)\sim \sqrt{\frac{2}{\pi}}(j_l(kr)\cos\delta -y_l(kr)\sin\delta),\quad r\rightarrow \infty,
  \label{eq:8}
\end{equation}
where $\dfrac{\hbar^2 k^2}{2\mu}=E_n$, $\delta$ is the phase shift, $j_l, y_l$ is the first and the second kind spherical bessel functions. By matching the obtained wave function of the scattering states and the asymptotic behaviour Eq. (\ref{eq:8}), the phase shift can be obtained as follows:
\begin{equation}
\tan\delta=\frac{kj'_l(kR)\psi_{nl}(R) - j_l(kR)\psi'_{nl}(R) }{ky'_l(kR)\psi_{nl}(R)-y_l(kR)\psi'_{nl}(R)},
\end{equation}
where $R$ is the match point which can take an arbitrary large enough value outside the range of the potential.
We take the neutron $d_{3/2}$ channel of $^{22}$O for example.
By taking the contour on the real axis, we obtain many scattering states with discretized energies. In Fig. \ref{fig:wavefun}, we show the wave function of the calculated state with an energy $E=50.7$ MeV as well as  the asymptotic wave function. The match point $R=10$ fm is taken and the phase shift that makes the two wave functions match is 1.5 rad. We can see the calculated wave function indeed behaves in coincidence with the asymptotic wave function at large distance.

\begin{figure}[]
\centering
\includegraphics[scale=1.0]{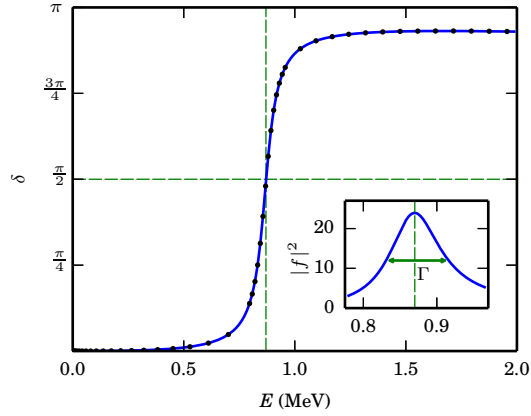}
\caption{The phase shift as a function of the scattering energy. The inset graph shows the norm square of the scattering amplitude as a function of the energy.}
\label{fig:phaseShift}
\end{figure}

Scattering states with different energies have different phase shifts. By analyzing the energy dependence of the phase shift, we can obtain the position of the resonance. On the other hand, with a complex contour, the resonance state in neutron $d_{3/2}$ channel of $^{22}$O has already been found at $E=0.867-0.042i$ in the previous subsection. We can check whether the two calculations are consistent. Figure \ref{fig:phaseShift} shows the relation of the phase shift against the energy. The rapid change of the phase shift across $\pi/2$ indicates the existence of a resonance state there. The energy $E_r$ at which gives a phase shift of $\pi/2$ is the center energy of the resonance. While the width of the resonance $\Gamma$ is given by
\begin{equation}
\frac{\Gamma}{2}=-\left.\frac{1}{d\cot\delta/dE}\right|_{E=E_r}=\left.\frac{1}{d\delta/dE}\right|_{E=E_r}.
\end{equation}
In Fig. \ref{fig:phaseShift}, to illustrate the position and the width of the resonance state, we also show the norm square of the scattering amplitude, which is linear related to the scattering cross section. The scattering amplitude reads
\begin{equation}
f=\frac{1}{k\cot\delta-ik}.
\end{equation}
The resonance extracted from the phase shift analysis is $0.87-0.04i$ MeV, which is consistent with the calculation in previous subsection.

\section{Summary}
We present a new simplified method to generate the Hartree-Fock Gamow basis from the realistic nuclear force. We first perform the HF iteration in HO basis, and then the HF potential obtained is analytically continued to the complex $k$-plane. The continuation is accomplished directly by a basis transformation in which the complex Laguerre polynomials are used. By discretizing the integral of the Schr\"odinger equation in momentum space, the equation becomes a complex symmetric eigenvalue problem. The Hartree-Fock Gamow basis can be obtained by diagonalizing the complex symmetric matrix. The method is tested for $^4$He and $^{22}$O with the renormalized N$^3$LO interactions. The basis obtained can be further used for studies of weakly bound nuclei.

\begin{acknowledgments}
  This work has been supported by the National Natural Science
  Foundation of China under Grants No. 11235001, No. 11320101004 and No.
  11575007; and the CUSTIPEN (China-U.S. Theory Institute for Physics with
  Exotic Nuclei) funded by the U.S. Department of Energy, Office of Science
  under Grant No. DE-SC0009971.
\end{acknowledgments}

\end{document}